\documentclass[%
 reprint,
 amsmath,amssymb,
 aps,
]{revtex4-2}
\usepackage{graphicx}
\usepackage{dcolumn}
\usepackage{bm}
\usepackage{euler}
\usepackage{subcaption}
\usepackage{xcolor}
\usepackage{amsmath}
\usepackage{amssymb}
\usepackage{parskip}

\begin{document}

\preprint{APS/123-QED}

\title{The Equation of State with the EPOS3 model}

\author{Maria Stefaniak}

\affiliation{%
 Department of Physics, The Ohio State University, 43210 Columbus, Ohio, USA \\
 \\
 GSI Helmholtzzentrum für Schwerionenforschung, Planckstr. 1, 64291 Darmstadt, Germany \\
}%

\author{Klaus Werner\\}
\affiliation{
SUBATECH, University of Nantes – IN2P3/CNRS – IMT Atlantique, Nantes, France
}%

\author{Hanna Paulina Zbroszczyk}
\affiliation{%
 Warsaw University of Technology, Faculty of Physics, Koszykowa 75, 00-662 Warsaw, Poland
}%
\author{Johannès Jahan}
\affiliation{%
Department of Physics, University of Houston, Houston, TX 77204, USA
}%
\date{\today}

\begin{abstract}

Transitions between different states of matter and their thermodynamic properties are described by the Equation of State (EoS). A universal representation of the EoS of Quantum Chromodynamics (QCD) for the wide range of phase diagram has yet to be determined. The expectation of the systems to undergo various types of transitions depending on the temperature (T), the chemical potential ($\mu_B$), and other thermodynamic features make solving that puzzle challenging. Furthermore, it needs to be apparent which experimentally measurable observables could provide helpful information for determining EoS. The application of different EoS for hydrodynamical evolution was introduced in the EPOS3 generator, which allows one to study its changing effect on the experimental observables. The family of EoS proposed by the BEST Collaboration was implemented. The Critical Point (CP) location and the strength of criticality variations were investigated with particle yield, transverse momentum spectra, flow, and moments of the net-proton distributions.

\end{abstract}

\maketitle

\section{\label{sec:Intro}Introduction}

Determining the EoS is crucial for the complete description and understanding of the QCD phase diagram. The relations between thermodynamic quantities characterizing different states of matter are depicted in the construction of EoS. The substantial topic of the present research is the investigation of transitions between partonic and hadronic mediums. Depending on the medium's T and $\mu_B$, it is expected to undergo smooth cross-over or rapid first-order phase transition. EoS expresses the relations between various matter parameters such as pressure, temperature, energy density, speed of sound, and the former. 
It is not trivial to determine it for the broad range of the $\mu_B$. At $\mu_B=0$ and extreme T, one can apply the non-perturbative QCD and based on the first principles Lattice QCD computations \cite{Allton:2002zi, Redlich:2004gp, Gavai:2008zr}. It provides quantitative information on the deconfined state QGP and \textit{cross-over} transition. Even though applying increasingly sophisticated algorithms \cite{Stephanov:2007fk}, the area of the figure at non-zero $\mu_{B}$ is still not fully understood. The existence and placement of the CP, where the cross-over transition switches to possible first-order phase one, cannot be predicted using fundamental principles.

Various attempts are performed to generate the EoS, allowing one to characterize the whole QCD phase diagram, starting from $\mu_{B} = 0$ and ending with higher baryon density matter. Some EoS introduce the first-order phase transition for finite $\mu_{B}$ and relatively lower T. Nevertheless, several provide information about the CP location and properties of this phase transition \cite{Stephanov:2007fk, Parotto:2018gic}. 

\section{\label{sec:besteos}BEST EoS}

Collaboration Beam Energy Scan Theory (BEST) proposed a family of EoS describing the same region of the QCD phase diagram as studied in the BES program \cite{Parotto:2018gic, Parotto:2018pwk}. BEST covered the region of $\mu_{B}$ in the range \mbox{$0$ - $450$ MeV} and T between \mbox{$30$ MeV - $800$ MeV}. The equations respect the lattice QCD results up to  \( \mathcal O \)($\mu_{B}^{4}$). They consider the existence of cross-over transition, and first-order phase transition and give a possibility to choose the location of CP on the QCD phase diagram $\{T, \mu_B\}$. The coverage of EoS at finite $\mu_B$ is possible due to the applied by BEST strategy \cite{Parotto:2018gic, Parotto:2018pwk}:
\begin{enumerate}
    \item Describe the universal scaling behavior of the EoS in the 3D Ising model close to the CP using an appropriate parametrization;
    \item The 3D Ising model phase diagram is mapped using a parametric change of variables onto the QCD one (Ising variables to QCD coordinates: $(h, r) \mapsto (T, \mu_{B})$), where $h$ is magnetic field, and $r$ is reduced temperature $r = (T-T_C)/T_C$. 
    \item Estimate the critical contribution to the expansion coefficients from Lattice QCD using the thermodynamics of the Ising model EoS;
    \item Reconstruct the full pressure, incorporating the proper critical behavior and matching Lattice QCD at $\mu_{B}=0$.
    
\end{enumerate}

\begin{figure*}
\centering
\includegraphics[width=0.7\textwidth]{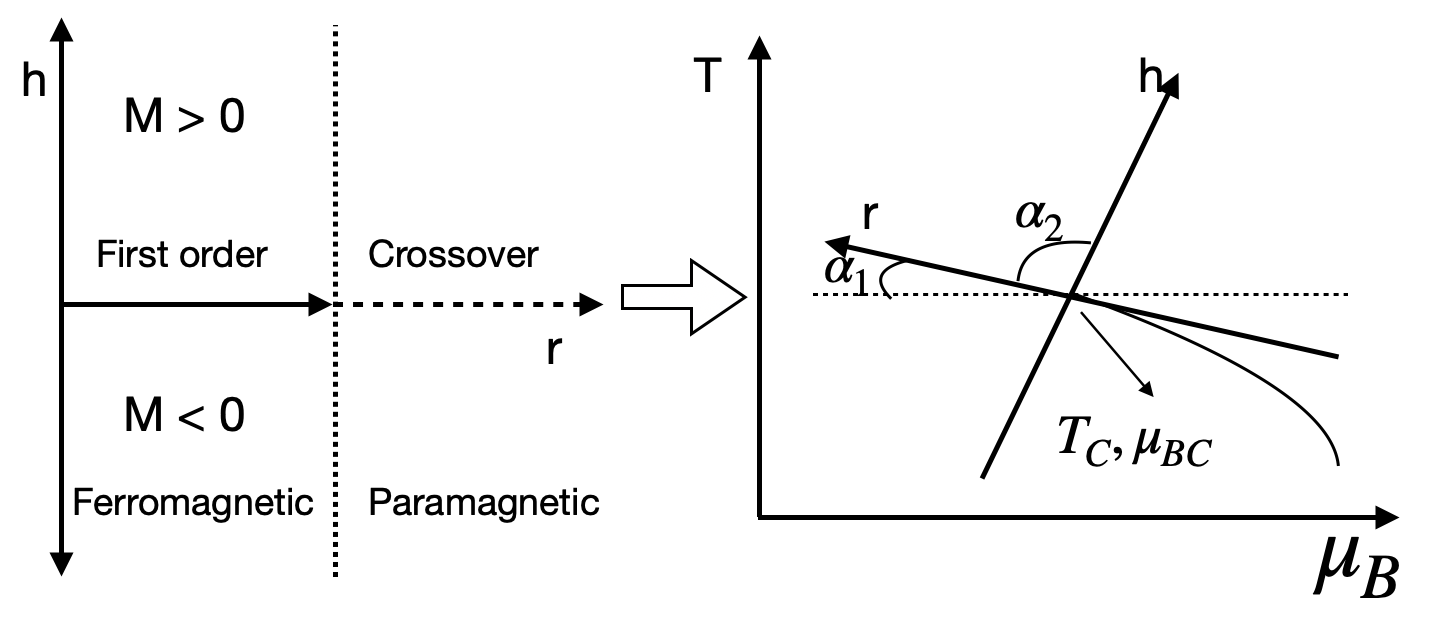}
\caption{The linear transformation is used in mapping the 3D Ising model diagram on the QCD one \cite{Parotto:2018pwk}.}
\label{fig:isingmap}
\end{figure*}

In Fig. ~\ref{fig:isingmap}, the non-universal mapping procedure is illustrated. Using six parameters (listed and described in Sec. \ref{sec:EoSPar}), the critical thermodynamics is linearly transferred to QCD.



\subsection{Selection of the EoS parameters}\label{sec:EoSPar}

The primary studies possible due to the modifications and development of the EPOS3 model \cite{Werner:2019mdw} allow one to study the impact on final observables of the changes between the variety of EoS. As mentioned in Section \ref{sec:besteos}, the BEST EoS is, in reality, the family of EoS, the set of various EoS tables. To obtain the EoS, one must choose the parameters corresponding to mapping properties and locate the CP on the QCD phase diagram. 

The composition of the input parameters is crucial in setting the strength of the criticality of the transitions of the matter. Moreover, by changing the CP's location to some extreme values, one can expect the cross-over or the first-order transitions in the evolution of examined simulated systems. The structure of the parameter input file is as follows:

\begin{equation*}
    MODE \quad T_0 \quad \kappa \quad \mu_{BC}  \quad \Delta_{\alpha_1, \alpha_2}  \quad\omega  \quad\rho 
\end{equation*}

Where (all visualized in the right panel of Fig. \ref{fig:isingmap}):
\begin{itemize}
    \item MODE - corresponds to way of locating the CP on the diagram. In this study, the CP lies on a parabola parallel to the chiral transition line - which reports to MODE = PAR;
    \item $T_0$ - the value of T at which the parabolic pseudo-critical line crosses the T axis;
    \item $\kappa$ -  the curvature of the transition line at the T axis;
    \item  $\mu_{BC}$ and $T_C$ -  the $\mu_{B}$ and $T$ at the CP;
    \item $\Delta_{\alpha_1, \alpha_2}$ - the difference between two angles shown in Fig. \ref{fig:isingmap}.
    \item $\omega$ - the global scaling parameter in the mapping (the higher $\omega$ the less criticality in transitions of matter);
    \item $\rho$ - the relative scaling in the mapping; both $\omega$ and $\rho$ application described in more details in \cite{Parotto:2018pwk};
\end{itemize}

The $T_C$, $\alpha_{1,2}$ can be easily calculated from the given parameters:
\begin{subequations}
\begin{eqnarray}\label{eq:TcIsing}
T_C = T_0 +\kappa/T_0 \mu_{BC}^{2}, \label{appa}
\\
\alpha_1 = 180/\pi \bigg| \arctan(-\frac{2\kappa}{T_0 \mu_{BC}})\bigg|, \label{appb}
\\
 \alpha_2 = \alpha_1 + \Delta_{\alpha_1,\alpha_2} \label{appc}
\end{eqnarray}
\end{subequations}

\section{\label{sec:EPOS}EPOS3 Model}
EPOS3 is an abbreviation of \textbf{E}nergy conserving quantum mechanical multiple scattering
approach, based on \textbf{P}artons (parton ladders), \textbf{O}ff-shell remnants,
and \textbf{S}aturation of parton ladders.

The model consists of several phases of evolution: 
\begin{itemize}
    \item initial stage (based on the Parton Gribov-Regge theory) \cite{Gribov:1968fc, Drescher:2000ha}, 
    \item core/corona division \cite{Werner:2013tya, Werner:2010aa, Werner:2007bf}
    \item hydrodynamical evolution \cite{Werner:2010aa},
    \item hadronization based on the given EoS,
    \item hadron rescattering  (based on UrQMD \cite{Bleicher:1999xi,Bass:1998ca}),
    \item resonance decays.
\end{itemize} 

The crucial element of the model's theoretical framework is the sophisticated treatment of the hadron-hadron scattering and the initial stage of the collisions at ultra-relativistic energies. It is highly relevant in the understanding of possible parton-hadron phase transition. The merged approach of Gribov-Regge Theory (GRT) and the eikonalised parton model is utilized to treat the first interactions happening just after the collision properly - satisfying conservation laws and equal treatment of subsequent Pomerons \cite{Drescher:2000ha, Gribov:1968fc, WERNER199387}.

If the density of the strings is very high, they cannot decay independently, what describes scenario of the heavy-ions and the high-multiplicity $pp$ collisions. In EPOS3, the dynamical process of division of the strings segments into \textit{core} and \textit{corona} is introduced in order to deal with this issue \cite{Werner:2013tya, Werner:2010aa, Werner:2007bf}.

The separation is based on the abilities of a given string segment to leave the "bulk matter". As the criteria for deciding if it goes to \textit{core} or \textit{corona}, the transverse momentum of the element and the local string density are considered. If the string segment belongs to the very dense area, it will not escape but will build the \textit{core}, which will be driven in the next step by a hydrodynamical evolution. When the segment originates from the part of the string close to a kink, characterized by the high transverse momenta, it escapes the bulk matter and joins the \textit{corona} and consequently will show up as a hadron (jet-hadrons). There is also a possibility that the string segment is close to the surface of the dense part of the medium, and its momentum is high enough to leave it; it also becomes a \textit{corona} particle. The following equation is used for the determination of the core and corona:
\begin{equation}
    p^{new}_{t} = p_{t} - f_{Eloss}\int_{\gamma} \rho dL
\end{equation}
where: $\gamma$ is the trajectory of the segment,  $\rho$ the string density, and $f_{Eloss}$ a non-zero constant for $p_T>p_{T,1}$, null for $p_T<p_{T,2}$ and interpolated linearly between $p_{T,1}$ and $p_{T,2}$. If the $p^{new}_{t}$ is positive for a given segment, it escapes and becomes a corona particle; in the opposite case, it contributes to the core.
\begin{figure}
\centering
\includegraphics[scale=0.5]{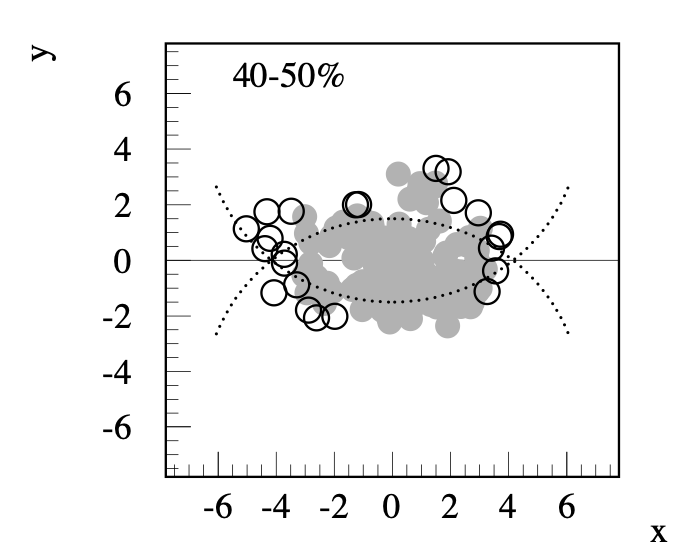}
\caption{The Monte Carlo simulation: full circles correspond to string segments contributing to core and open to corona. The big circles are just for the eye guidance showing the surface of collided nuclei \cite{Werner:2007bf}}
\label{fig:string_seg}
\end{figure}

As it has been studied \cite{PhysRevLett.102.172302,PhysRevC.79.044914,Snellings:2011sz,Gale:2013da}, the QGP does not expand like an ideal fluid, and the effect of the bulk viscosity has to be taken into account in the simulations. In EPOS3, the 3D+1 viscous hydrodynamics is applied, providing an proper description of the collective expansion of the matter \cite{Werner:2010aa}. 
The hydrodynamic evolution is based on the EoS. In this project we introduced the possibility to change EoS and apply the BEST ones.

In the simulations, the definitive treatment of individual events is essential - the generalization in considering smooth initial conditions for all events is not applied. 
The event-by-event (\textit{ebe}) approach in hydrodynamical evolution is based on the random flux tube initial conditions \cite{Werner:2010aa}. It has a relevant impact on the final observables, such as spectra or various harmonics of flow. The hadronization process occurs according to the microcanonical approach described in \cite{Werner:1995mx,Klaus_micro}.

The final part of the simulation uses a so-called \textit{hadronic afterburner} - UrQMD \cite{Bleicher:1999xi,Bass:1998ca}. 

When the system's density is very high and the mean free paths of constituent particles are small about any macroscopic length scale, the hydrodynamic description can be used - in the initial phase of the QGP evolution. With the system's cooling, the density and the mean free paths decrease; oppositely, the $\eta / s$ increases. Finally, the differences in the mean free path of various particle species become relevant, and the system's collective description becomes inadequate. When the density and the temperature are low enough, the kinetic theory is applied using the UrQMD code \cite{Bleicher:1999xi,Bass:1998ca}. 

The particles can interact only when they leave the hyper-surface of the freeze-out. The $2\rightarrow n$ hadronic scattering is performed according to the measure reaction cross sections \cite{Patrignani:2016xqp}. Of the 60 different baryonic species and their antiparticles, about 40 mesonic states are considered \cite{Bass:1998ca,Bleicher:1999xi}. There are implemented such interactions between hadrons as \cite{Steinheimer:2018fja}:
\begin{itemize}
    \item elastic scattering,
    \item string excitations,
    \item resonance excitations,
    \item strangeness exchange reactions 
\end{itemize}
The hadronic scattering significantly impacts the final observables \cite{Stefaniak:2018wwh, Steinheimer:2016cir}.


\section{\label{sec:Results}Results and discussion}

\subsection{Simulations}

The two collision energies were studied: Au+Au collisions at \mbox{$\sqrt{s_{NN}} = 7.7$ GeV} and $27$ GeV. Below the \mbox{$\sqrt{s_{NN}} = 11.5$ GeV} the onset of QGP is expected according to STAR experimental results \cite{Pandit:2012mq, Adamczyk:2015fum, Adamczyk:2013gw}, which motivates the choice of the lower collision energy. The second one is the medium one in BES-I at RHIC \cite{Odyniec:2013kna}.
The EPOS3 model simulations where performed using following EoS:
\begin{itemize}
    \item X3F cross-over, 3 flavour conservation \cite{Werner:2013tya}
    \item BEST EoS with various parameters listed in \mbox{Table \ref{tab:eos_par_tab}}.
\end{itemize}

The substantial statistics is needed for the precise studies of narrow centrality binning. In this research, we performed the preliminary investigation using lower number of events but looking into effects of various EoS. 

\begin{table*}[ht]
\caption{\label{tab:eos_par_tab} sets input parameters for constructing nine BEST EoS used in the EPOS3 simulations. Left part corresponds to input parameters for the BEST EoS construction code, on the right columns include calculated output variables. }
\centering
\begin{ruledtabular}
\begin{tabular}{l| l l l l l l l |l l l l l l}
\hline\hline
 Number  &  $MODE$  & $T_0 $ & $\kappa $ & $\mu_{BC}  $ & $ \Delta_{\alpha_{1,2}}  $ & $\omega  $ &  $\rho $ & $T_C$ & $\mu_{BC} $ & $\alpha_1 $ & $ \alpha_2   $ & $\omega T_C  $ & $\rho \omega T_C  $
 \\ [0.5ex] 
\hline
\toprule
BEST 1:   & PAR& 155 &-0.0149& 350 &90 &1& 2    &  143 &350 &3& 93 &143 &286\\
 \hline
BEST 2:   & PAR &155& -0.0149& 350 &90& 4& 1    &  143& 350 &3& 93& 572& 572\\
 \hline
BEST 3:   & PAR& 155& -0.0149& 420& 90& 0.75 &2 & 138& 420& 4& 94& 103& 207\\
 \hline
BEST 4:   & PAR &155 &-0.0149 &350 &90 &10 &1   & 143& 350& 3 &93& 1432& 1432\\
 \hline
BEST 5:   & PAR &169 &-0.0149& 420& 90& 1& 1    &  153& 420& 4 &94 &153& 153\\
 \hline
BEST 6:   & PAR &169& -0.0149& 420& 90& 0.5 &1  & 153 &420& 4 &94 &76 &76\\
 \hline
BEST 7:   & PAR &174& -0.0149 &440& 90 &1 &1    & 157 &440 &4 &94 &157 &157\\
 \hline
BEST 8:   & PAR &178 &-0.0149& 300& 90& 1 &1    & 170 &300& 2 &92& 170 &170

\end{tabular}
\end{ruledtabular}
\end{table*}

\begin{figure}[!ht]
\centering
\includegraphics[width=0.4\textwidth]{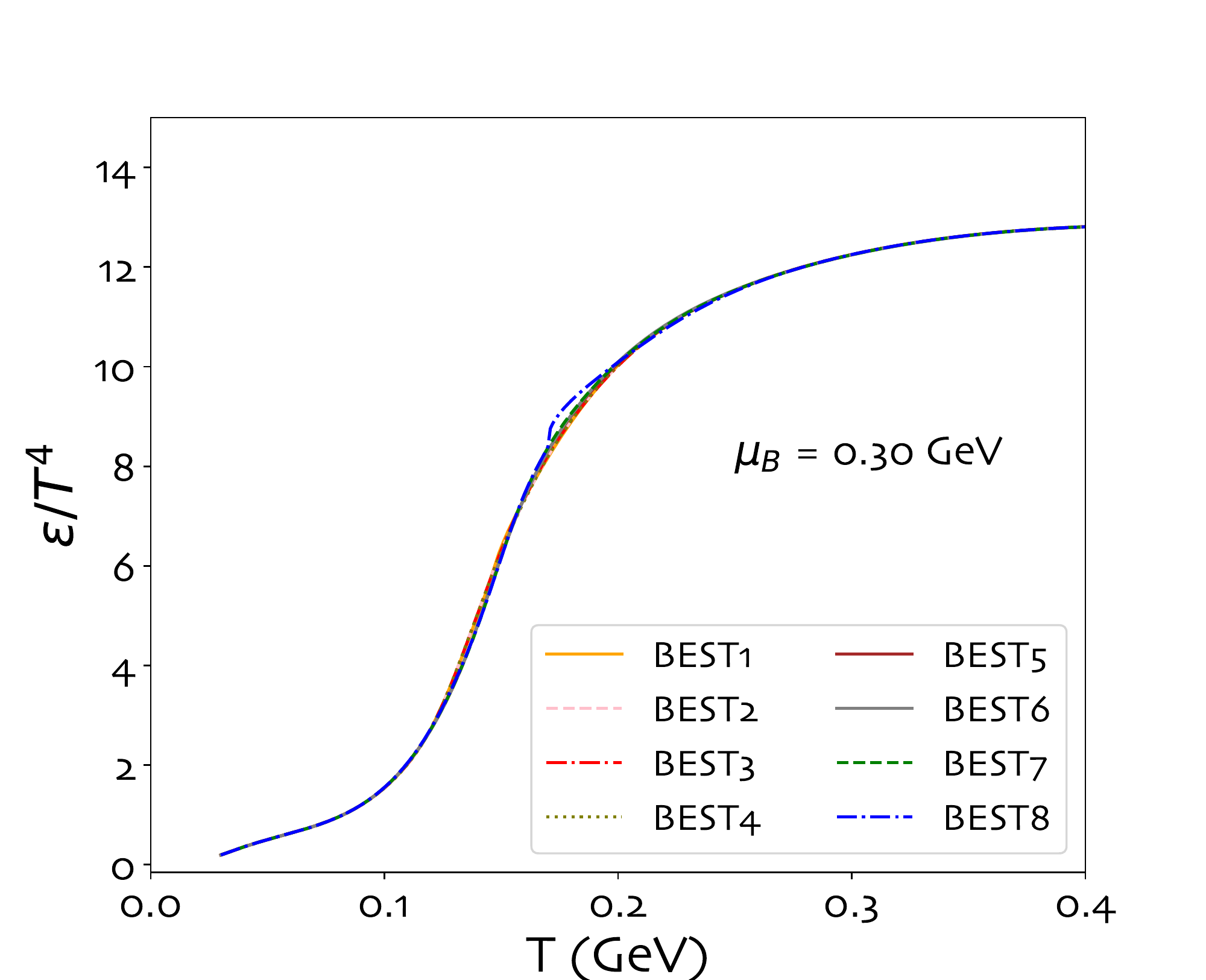} 
 \caption{Energy density as a function of temperature at $\mu_B$ = 300 MeV for all constructed EoS.}
\label{fig:2deosPlot_eng1}
\end{figure}

\begin{figure}[!ht]
\centering
\includegraphics[width=0.4\textwidth]{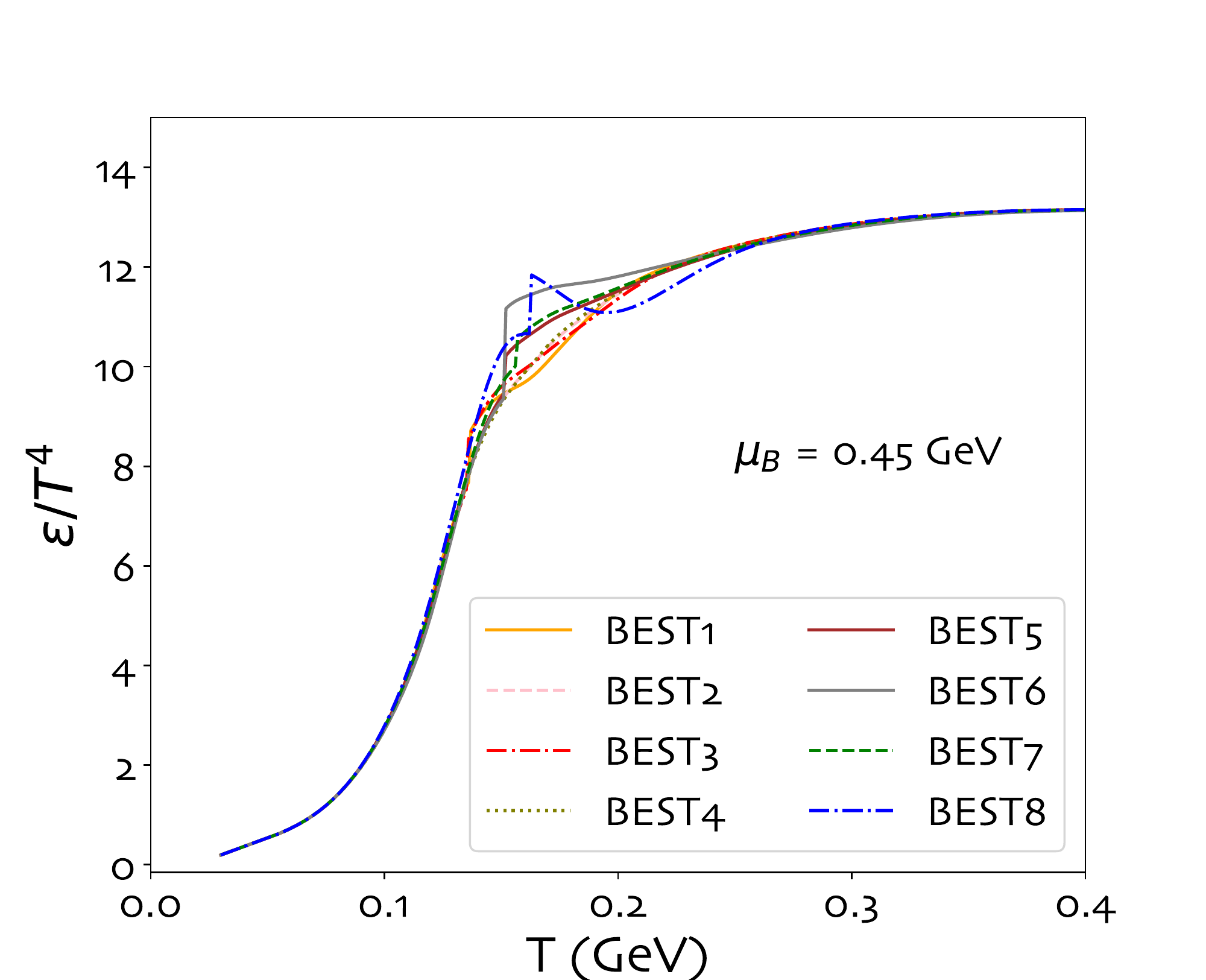}  
 \caption{Energy density as a function of temperature at $\mu_B$ = 450 MeV for all constructed EoS.}
\label{fig:2deosPlot_eng2}
\end{figure}

\begin{figure}[!ht]
\centering
\includegraphics[width=0.4\textwidth]{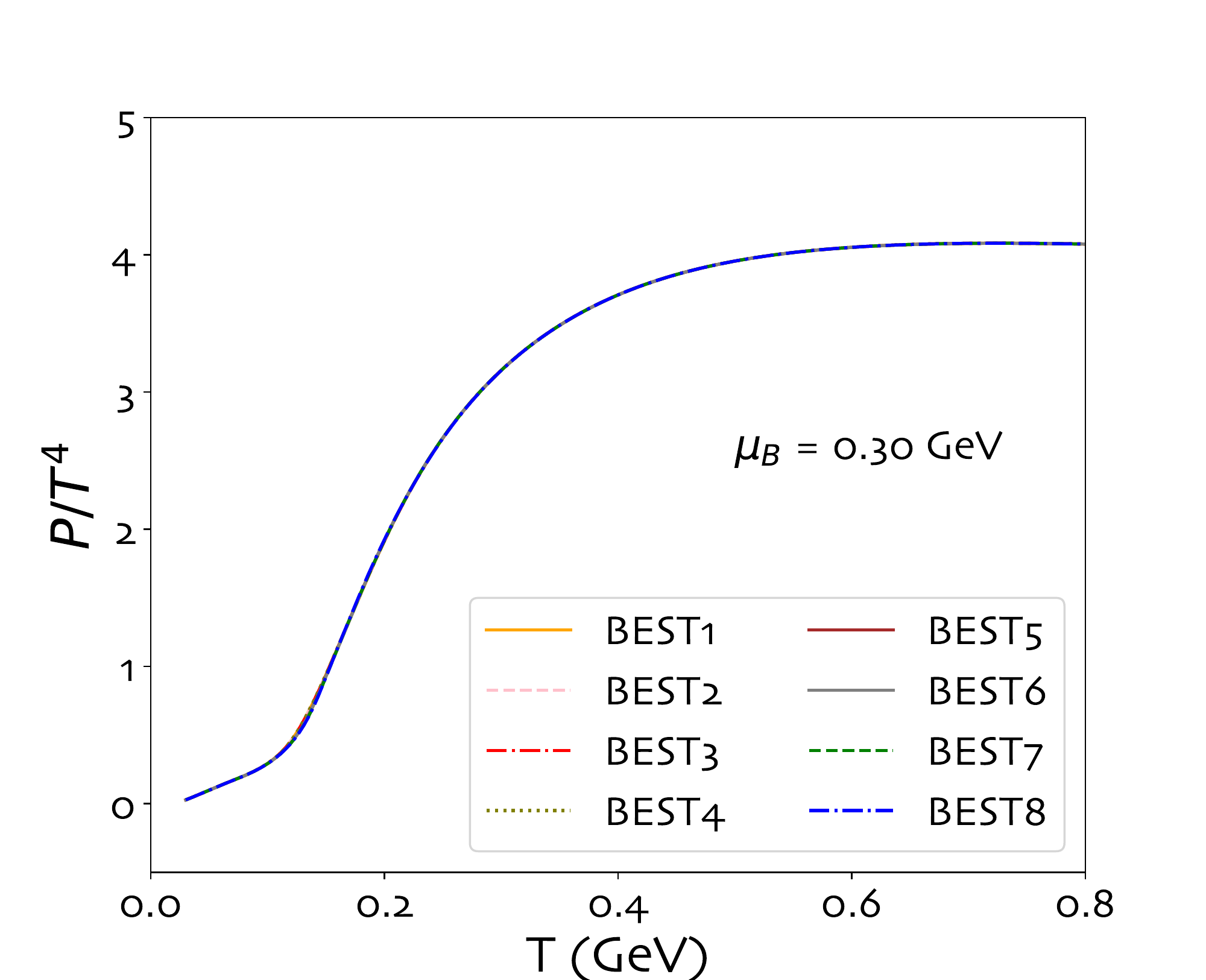} 
 \caption{Pressure as a function of temperature at $\mu_B$ = 300 MeV for all constructed EoS.}
\label{fig:2deosPlot_p1}
\end{figure}

\begin{figure}[!ht]
\centering
\includegraphics[width=0.4\textwidth]{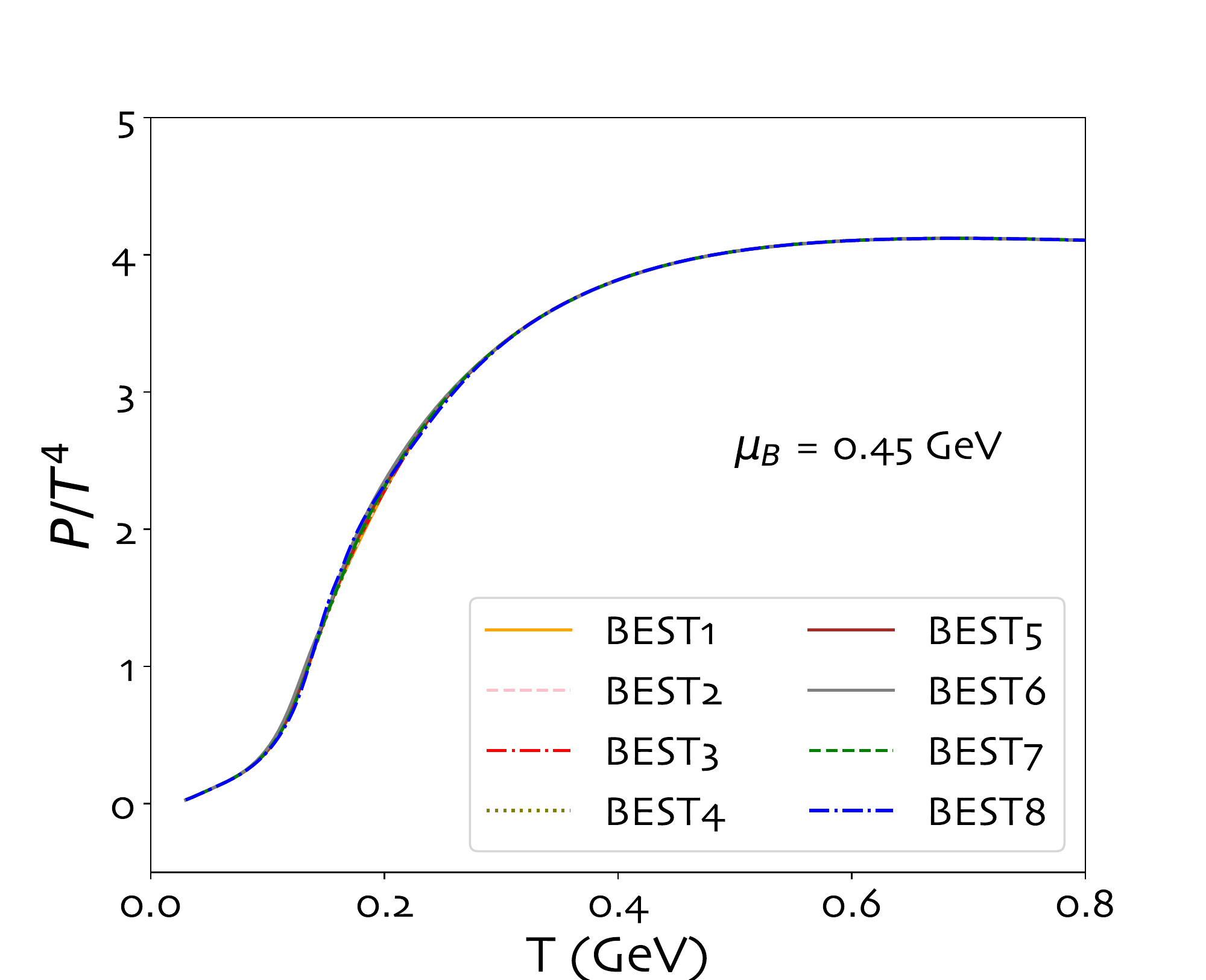}  
 \caption{Pressure as a function of temperature at $\mu_B$ = 450 MeV for all constructed EoS.}
\label{fig:2deosPlot_p2}
\end{figure}

The only element of the simulation at given energy performed by the EPOS3 model which changes is the EoS. All the presented data sets can be used to directly compare the proposed EoS. The Figures \ref{fig:2deosPlot_eng1}-\ref{fig:2deosPlot_p2} illustrate the dependencies of energy density and pressure with temperature for each EoS at \mbox{$\mu_B$ $0.30$ GeV} and \mbox{$0.45$ GeV.} The significant variations due to the presence of CP are visible in the density energy plots in the $T = 0.12 - 0.2$ GeV region.

\subsection{Production of particles}
\begin{figure}[!ht]
\centering
\includegraphics[width=0.4\textwidth]{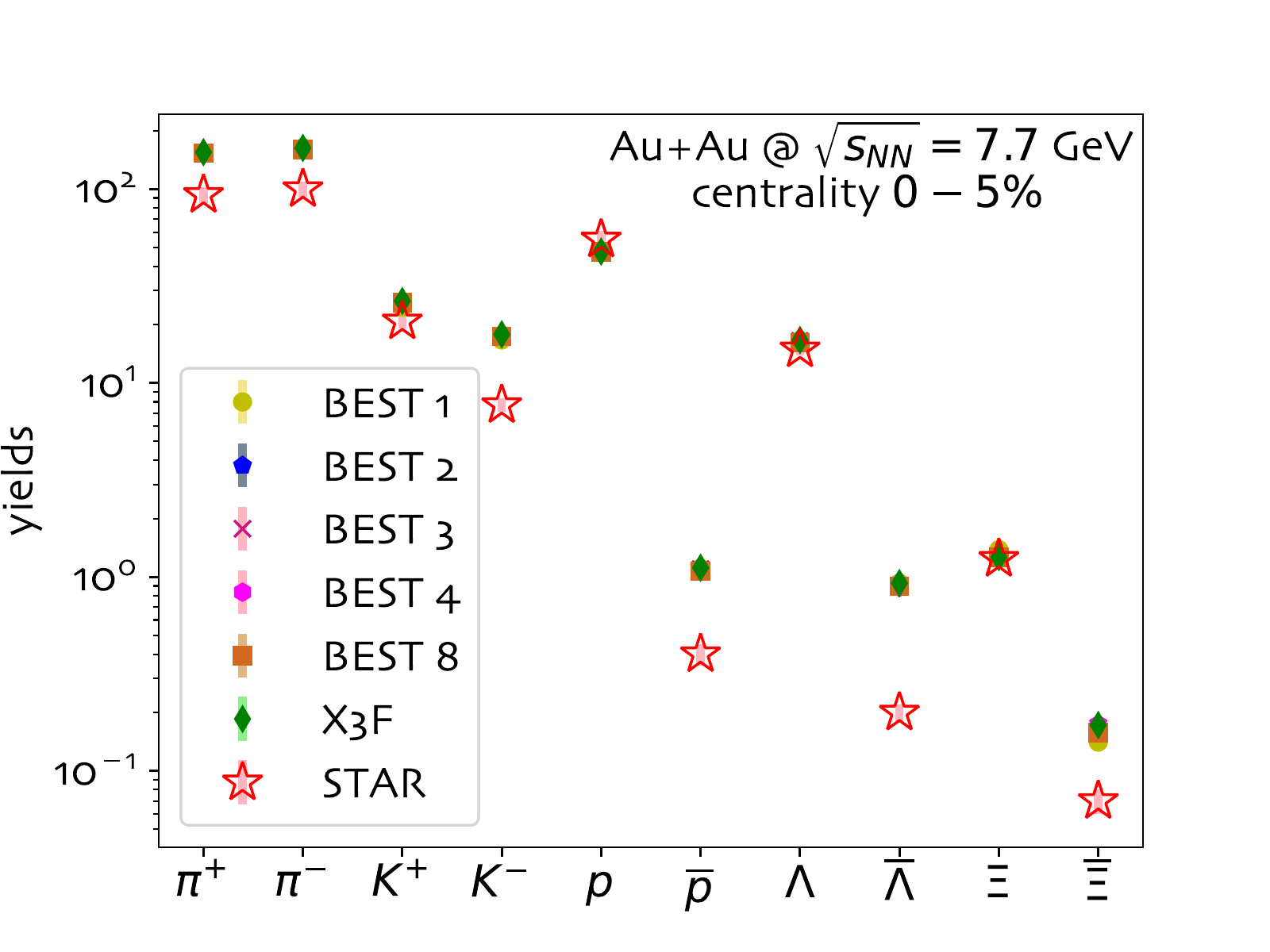} 
 \caption{Particle yields for Au+Au most central $0-5\%$ collisions at \mbox{$\sqrt{s_{NN}} = 7.7$ GeV} simulated with EPOS3 model using various EoS and compared with STAR data \cite{Adamczyk:2017iwn}.}
\label{fig:yields_epos}
\end{figure}

\begin{figure}[!ht]
\centering
\includegraphics[width=0.4\textwidth]{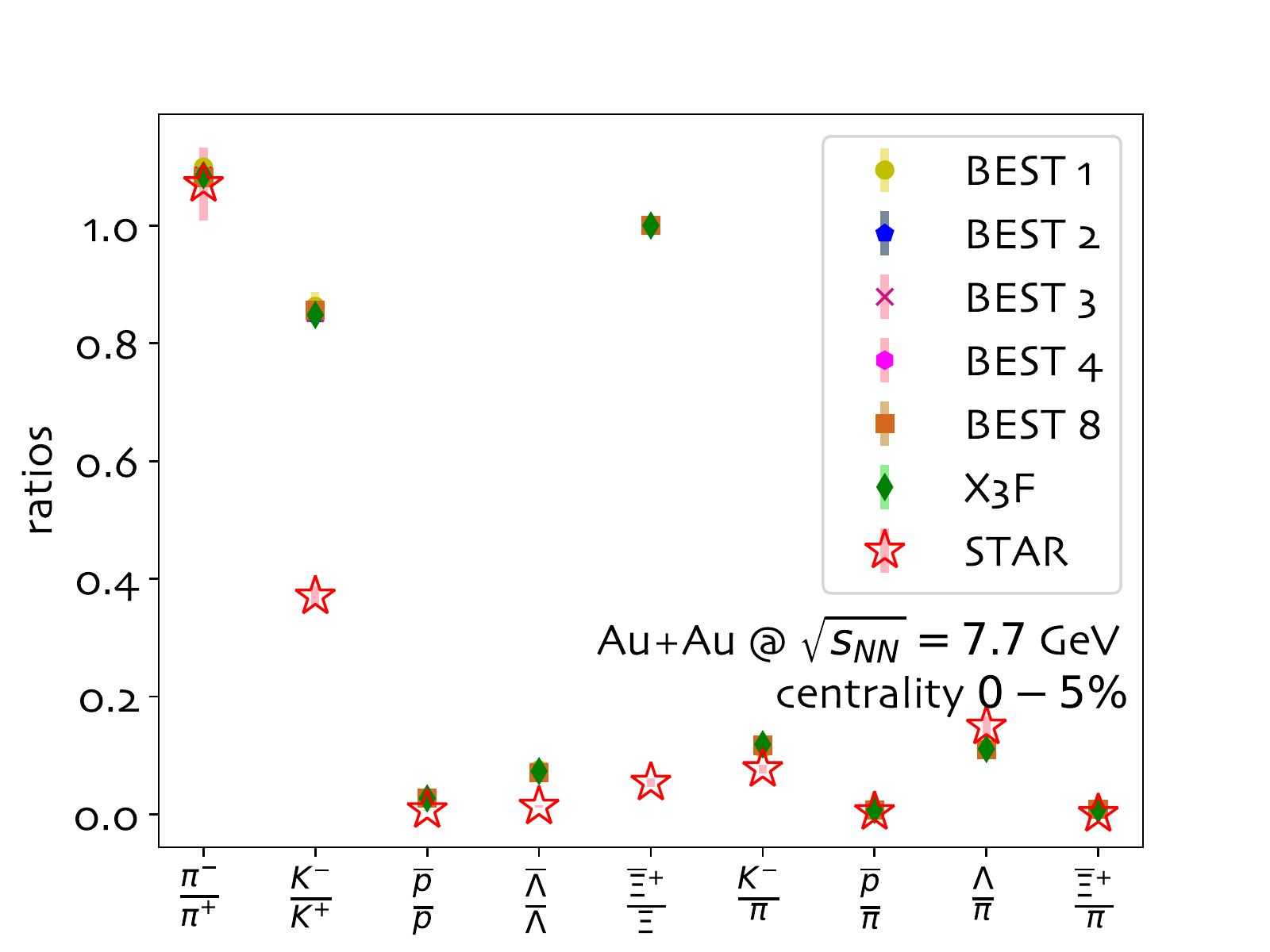}  
 \caption{Particle ratios for Au+Au most central $0-5\%$ collisions at \mbox{$\sqrt{s_{NN}} = 7.7$ GeV} simulated with EPOS3 model using various EoS and compared with STAR data \cite{Adamczyk:2017iwn}.}
\label{fig:ratios_epos}
\end{figure}

 Figure \ref{fig:yields_epos} shows the particle production at the most central $0-5\%$ collisions of Au+Au simulated with EPOS3 model. They are compared with STAR data published in \cite{Adamczyk:2017iwn}. Centrality in the model are defined using the Glauber model. Various EoS sets of parameters were used in performed simulations; the numbers of EPOS3 data sets correspond to those listed in Table \ref{tab:eos_par_tab}. As all the points from simulations precisely overlap each other, so only part of the data sets were plotted. The relations between particles' and antiparticles' production are reflected using ratios in Figure \ref{fig:ratios_epos}. 

The higher number of produced baryons than antibaryons proves that in EPOS3 simulations, the impact of non-zero baryon potential is kept for all the proposed EoS. The model reflects the experimental data reasonably, except for pions twice overestimated. 
Nonetheless, their ratio is kept. 

The possible reason for such discrepancies is the too-wide rapidity distribution of simulated data. In the experimental analysis, the selection of particles characterized by the $|y|< 0.1$ is very narrow. In such a case, even a tiny deviation in the rapidity distribution strongly affects the performed investigation. 

 In both Fig. \ref{fig:yields_epos} and \ref{fig:ratios_epos}, no relevant differences between simulations obtained with various EoS are observed. Notwithstanding the $X3F$ EoS corresponds to the cross-over transition, which is not expected to happen for cooling systems created in collisions of Au+Au at \mbox{$\sqrt{s_{NN}} = 7.7$ GeV}.

\subsection{Particles' dynamics}

The differences between the EoS were searched in the dynamics of the expanding matter. The listed below observables were investigated:
\begin{itemize}
    \item transverse momentum ($p_T$) spectra of identified hadrons: p, $\bar{p}$, $K^{\pm}$, $\pi^{\pm}$ (Au+Au at $\sqrt{s_{NN}} = 7.7$ and $27$ GeV, $0-5\%$ and $60-80\%$ centrality ranges),
    \item elliptic flow ($v_2$) of identified hadrons: p, $\bar{p}$, $K^{\pm}$, $\pi^{\pm}$ (Au+Au at $\sqrt{s_{NN}} = 7.7$ and $27$ GeV, $0-80\%$ centrality),
\end{itemize}

\begin{figure}[!ht]
    \centering
    \includegraphics[width=0.48\textwidth]{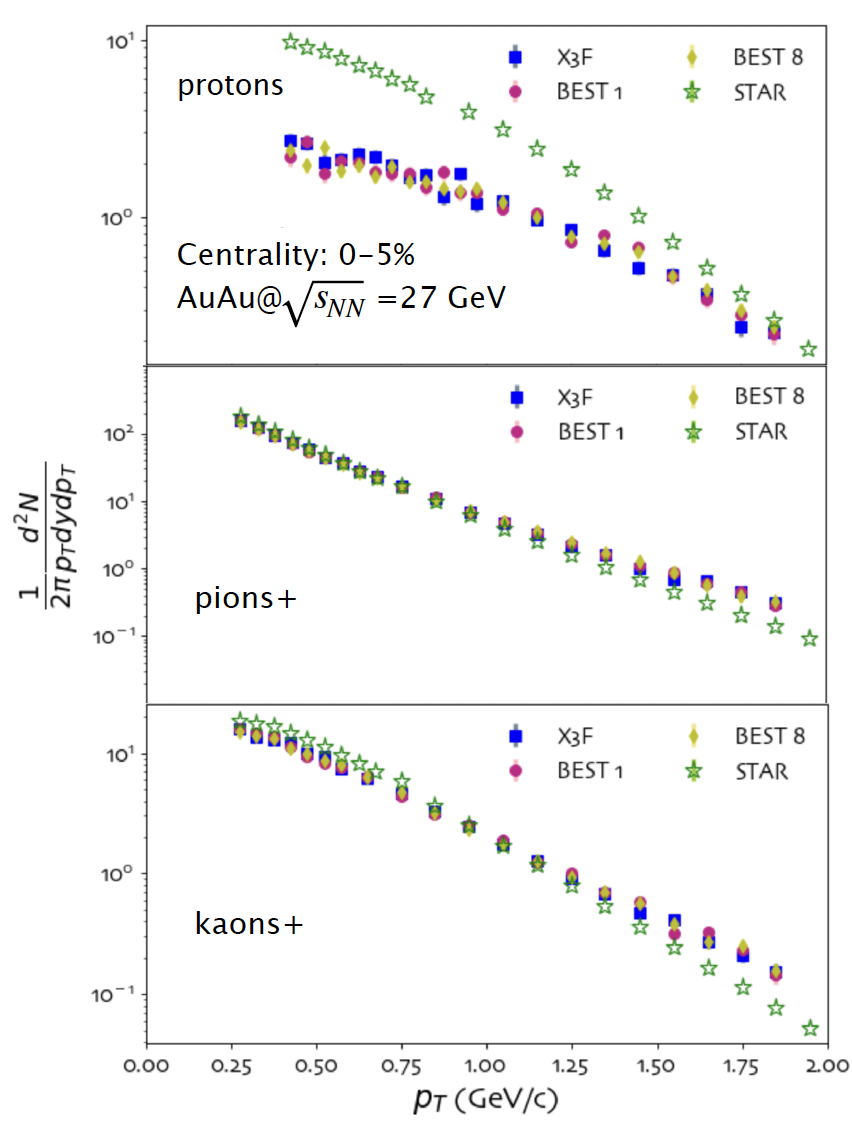}
    \caption{$p_T$ spectra of identified hadrons for Au+Au collisions at $\sqrt{s_{NN}} = 27 $ GeV. Simulated data compared with STAR experimental data \cite{Adamczyk:2017iwn}. }
    \label{fig:pt}
\end{figure}
\begin{figure}[!ht]
    \centering
    \includegraphics[width=0.48\textwidth]{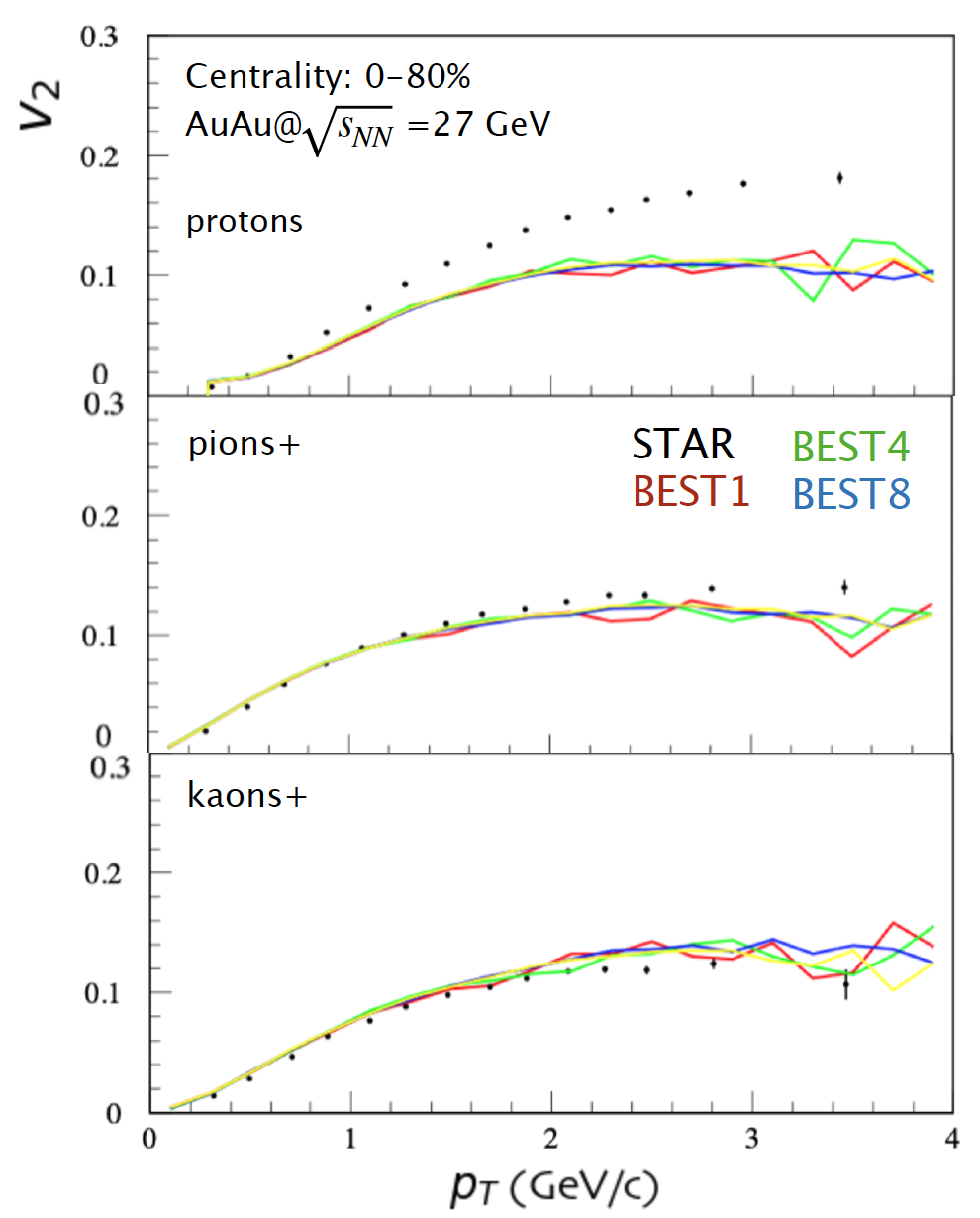}
    \caption{$v_2$ of identified hadrons for Au+Au collisions at $\sqrt{s_{NN}} = 27 $ GeV. Simulated data is shown with the curves and STAR experimental data \cite{Adamczyk:2017iwn} with black points. }
    \label{fig:flow}
\end{figure}

Surprisingly, none of the abovementioned observables depended on the applied EoS. Figures \ref{fig:pt} and \ref{fig:flow} show the comparison of the simulated EPOS3 $p_T$ and $v2$ with STAR experiment for Au+Au collisions at $\sqrt{s_{NN}} = 27$ GeV. 
Whereas the model describes the meson data reasonably well, there is clearly a problem with protons (which will be addressed in the future).

\subsection{Moments of particle distributions}

The non-monotonic behavior in the event-by-event fluctuations of globally conserved quantities is treated as one of the signatures of the presence of CP \cite{Sombun:2017bxi,Luo:2017faz}. The moments of distributions characterizing the given fluctuations are:  mean ($M$), standard deviation ($\sigma$), skewness ($S$), the kurtosis ($\kappa$).
They are linked with the corresponding higher-order thermodynamic susceptibilities and the system's correlation length \cite{Stephanov:2008qz,Athanasiou:2010kw}, which are expected to fluctuate for large samples in equilibrium at the CP. In the vicinity of CP, in reality, the system is driven away from the thermodynamic equilibrium, and the maximum value of correlation length attains 1.5-3 fm \cite{Athanasiou:2010kw}. During the fireball evolution after the hadronization stage, the freeze-out signal information can dissipate \cite{Stephanov:2009ra}. However, if it survives, the higher moments can become helpful in studies of CP's location. 
As the CP's location is changed in various EoS, the moments of particle distributions are expected to be a useful tool in the performed investigation. In the EPOS3 model, the critical fluctuations are not propagated in the hydro framework. However, still, the variations between different EoS could be visible.

\begin{figure}[!ht]
    \centering
    \includegraphics[width=0.4\textwidth]{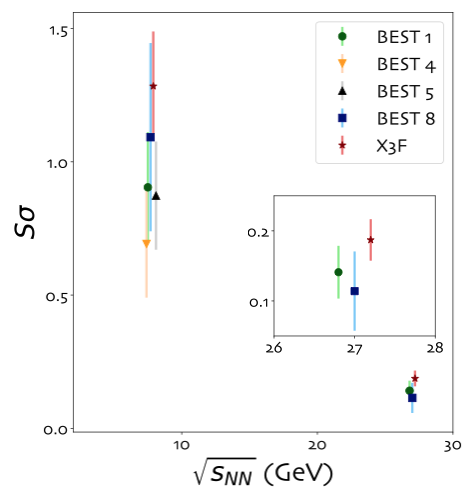}
    \caption{The $S\sigma$ of net-proton distributions for Au+Au collisions at $\sqrt{s_{NN}} = 7.7$ and \mbox{$27$ GeV } as a function of the collision energy $\sqrt{s_{NN}}$. The zoomed window corresponds to collisions at \mbox{$\sqrt{s_{NN}} = 27$ GeV}. The shifts of the Point on the x-axis are applied for better visualization.}
    \label{fig:ebe_momentsSNN1}
\end{figure}

\begin{figure}[!ht]
    \centering
    \includegraphics[width=0.4\textwidth]{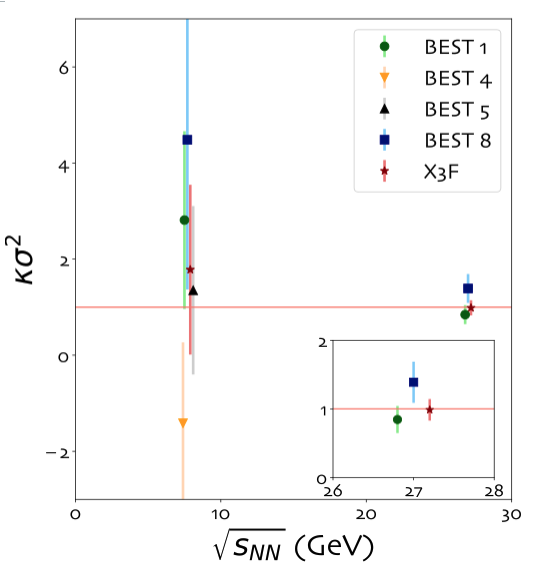}
    \caption{The $\kappa \sigma^2$ of net-proton distributions for Au+Au collisions at $\sqrt{s_{NN}} = 7.7$ and \mbox{$27$ GeV } as a function of the collision energy $\sqrt{s_{NN}}$. The zoomed window corresponds to  collisions at \mbox{$\sqrt{s_{NN}} = 27$ GeV}. The shifts of the Point on the x-axis are applied for better visualization.}
    \label{fig:ebe_momentsSNN2}
\end{figure}

Figures \ref{fig:ebe_momentsSNN1} and \ref{fig:ebe_momentsSNN2} show the $S\sigma$ and $\kappa \sigma^2$ integrated overall $N_{part}$ as a function of the collision energy. To perform this analysis in smaller centrality bins, enormous statistics are required. However, even in integrated data, significant energy dependence is present for $S\sigma$ for all EoS. All the points at the given energy are within the statistic uncertainties; effectively, no clear statement about the discrepancies between the EoS. $\kappa \sigma^2$ shows more considerable variations between different EoS data sets. At \mbox{$\sqrt{s_{NN}} = 7.7$ GeV}, the highest point corresponds to the EoS where the CP is located at high T and low $\mu_B$ and the simulated system is expected to go through the first-order transition. At the same time, the negative value is related to BEST4, where the criticality is less pronounced. For data sets simulated at \mbox{$\sqrt{s_{NN}} = 27$ GeV}, the differences are minor; however, the BEST8 value is the highest. The energy dependence is not definite.


The measurements of the net-proton distributions' moments show the differences between data simulated using various EoS. They are more pronounced in peripheral collisions where we do not expect an immense contribution from the \textit{core} particles, consequently, less dependent on the EoS.

\section{\label{sec:Conclusion}Conclusion}

The studies of various EoS implemented in the EPOS3 model were described. Developing the generator's code by introducing a new EoS gave a possibility to investigate the impact of EoS on the final observables. Apart from the EoS, the whole structure of the model remained unchanged.
EPOS3 model did not show the variations between different implemented EoS for most of the examined observables (like flow or yields). 
It concludes that the EPOS3 model is not sensitive to switching the EoS used in simulations. This version of the model is still under development. Studies based on higher statistics will be performed on the final model version, EPOS4. 

\section{\label{sec:Acknowledgments}Acknowledgements}

We thank Yurii Karpenko and Gabriel Sophys for the fruitful discussions.
This work was supported by the Grant of the National Science Centre, Poland, No: \vbox{2021/41/B/ST2/02409 and 2020/38/E/ST2/00019}. Studies were funded by IDUB-POB-FWEiTE-3, a project granted by Warsaw University of Technology under the program Excellence Initiative: Research University (ID-UB), Deutsche Akademische Austauschdienst, GET$\_$INvolved Programme,
and Humboldt-Forschungsstipendium für Postdocs, and  U.S. Department of Energy grant DE-SC0020651.
\bibliography{apssamp}


\end{document}